\documentclass{iaus260}
\usepackage{graphicx}
\pubyear{2009}
\volume{260}  
\pagerange{10--17}
\jname{The R\^ole of Astronomy in Society and Culture}
\editors{D. Valls-Gabaud \& A. Boksenberg, eds.}

\title[Fate of the Universe]                   
{The Big Bang, Modern Cosmology and the Fate of the Universe: Impacts upon Culture}  

\author[Lawrence M. Krauss]           
{Lawrence M. Krauss}        

\affiliation{School of Earth and Space Exploration and Department of Physics \\ 
                 Arizona State University, PO Box 871404, Tempe AZ 85287-1404\\ 
                 email: {\tt krauss@asu.edu} \\[\affilskip]
           }

\begin{document}
\maketitle

\begin{abstract}
Cosmological discoveries over the past century have completely changed our picture of our place in the universe.  New observations have a realistic chance of probing nature on heretofore unimaginable scales, and as a result are changing the nature of fundamental science.  Perhaps no other domain of science has an equal capacity to completely change our perspective of the world in which we live. 
\keywords{Big Bang, Cosmology, Culture.}    
\end{abstract}


Ever since humans have had the capacity to wonder, they have been inspired by the night sky to wonder about questions such as:  Where did we come from?, or,  Are we Alone?, or How will the Universe end?   These questions and others form the very basis of much of human culture, beginning with myths and religion, and moving, over the past 400 years into the domain of modern science.

If you stare up at the night sky from a dark place in the country, it is beautiful, to be sure.  However, we have realized over the past quarter century that the beauty we see is only the tip of a vast cosmic iceberg.  The really important stuff in the universe, at least the stuff that governed the formation of all cosmic structures and that will determine the ultimate fate of the universe itself, lies in the stuff we cannot see.  Over 95 $\%$ of the energy of the universe is invisible to telescopes, in the form of dark matter, which dominates the mass of essentially all galaxies, and even stranger, dark energy, which seems to permeate empty space, and accounts for almost 70 $\%$ of the energy in the universe.  Determining the nature of these two mysteries forms of mass and energy represent perhaps the greatest challenges to modern physics and cosmology.

As we ponder esoterica like dark matter and dark energy it is easy to lose sight of how much our picture of the universe has changed in even the span of a single human lifetime.    Eighty five years ago our universe consisted of a single galaxy, our Milky Way.  Now we know there are over 400 billion galaxies in the observable universe! 

More remarkable still, almost all people now take for granted the fact that our universe had a beginning which occurred a finite time ago. Literate people understand that beginning to have occurred in the Big Bang Explosion, some 13.7 billion years ago. (Alas, in the United States some 50 $\%$ of the US public still think the Universe is less than 10,000 years old, but I don't consider these people literate.)   But it is worth remembering that in 1916, after he had made his greatest theoretical discovery in developing the General Theory of Relativity, Albert Einstein, like essentially the entire scientific community, still thought we lived in a static eternal, and largely empty universe.  The fact that General Relativity does not allow a static solution with an extended mass distribution led Einstein to introduce his now famous Cosmological Constant as a fudge factor that he hoped might make his theory consistent with observations.  After Edwin Hubble unambiguously demonstrated a decade or so later that our universe is expanding, obviating the need for a cosmological constant,  Einstein called introducing it his biggest blunder.  More important than changing Einstein's views, an expanding universe implies that earlier on it was smaller, and earlier still smaller, until ultimately everything was together at a single point.

The Big Bang changed everything. So much so that the notion that the Universe had a beginning is now taken for granted across almost all cultures.  Interestingly the Big Bang has been seen on the one hand as a validation of religious notions of creation, and on the other as a direct challenge.  Most modern efforts to rid public school classes of the basis of modern biology, namely evolution, also try and remove discussions of the Big Bang.  

It was not always that way.  When the Belgian priest and physicist Georges Lema"tre first demonstrated mathematically, in 1931--shortly after Hubble's discovery-- that the equations of general relativity actually required a big bang--he was hailed by many, including Pope Pius XII, who 20 years later claimed that Lema"tre had proved Genesis. 

Incidentally, Lema"tre himself felt quite differently.  He initially inserted, then ultimately removed, a paragraph in the draft of his 1931 paper remarking on the possible theological consequences of his discovery. In the end, he said,``As far as I can see, such a theory remains entirely outside of any metaphysical or religious question."

The Big Bang is not a metaphysical theory, but a scientific one: namely one that derives from equations that have been measured to describe the universe, and that makes predictions that one can test.  Nevertheless, it is, to me, a remarkable example of how science can affect culture.  The very notion that the Universe had a beginning, whether or not that beginning is shrouded in mystery, is an essential part of the gestalt of modern life.  We should celebrate that. 

In the past decade there has been another discovery that will, I believe, ultimately have an equal impact upon our picture of our place in the cosmos, not because it affects our understanding of the past, but because it affects our picture of the future.  The discovery of dark energy has also changed everything about cosmology.  Before I explain why, I should quickly summarize the observational steps that inexorably led us to the strange conclusion that empty space contains most of the energy of the universe, and that that energy is gravitationally repulsive.


{\bf{\it Weighing the Universe}}. Every since Johannes Kepler derived his laws of planetary motion, leading Newton develop his Universal Law of Gravity, we have used gravity to weigh astronomical bodies, from the Earth and Sun, to the galaxy.  In order to determine the total matter content associated with the largest bound objects in the Universe, clusters of galaxies perhaps 10 million light years across, we have used not Newtonian Gravity, but General Relativity.  Einstein predicted in the 1930's, though he also suggested it would never be observed, that a sufficiently large concentration of mass could bend light coming from objects behind it, and in the process act like a ``gravitational lens", bending and distorting the image, magnifying it, or splitting it into many separate pieces.  Remarkably in the past two decades we have been able to detect these effects directly, visible in beautiful Hubble Space Telescope images for example.   Since we understand General Relativity, we can work backwards and determine both the distribution and amount of mass that would be required to produce the observed images.   This technique, among others has allowed us to weigh galaxy clusters, and derive a total accounting of the mass associated with all observed large scale structures in the universe.  The result was surprising, especially to those of us theorists who had long expected that we lived in a spatially flat universe today.  If one defines the quantity $\Omega$ as the ratio of the total density of matter in the universe divided by the density of matter required to result in an exactly flat universe today (the boundary between an open and closed universe), these observations (\cite{Tyson}) yielded:

\vskip 0.1in
\begin{equation}
\Omega_{m} \approx 0.30 \pm 0.1 \ (95\%)
\end{equation}
\vskip 0.1in

{\bf {\it Measuring the Geometry of the Universe Directly}}.  Consider the following question.  How would you measure the curvature of the Earth if you couldn't go around it, and couldn't see it from the outside?  Simple. Draw a large triangle.  On a flat Earth, the sum of the angles will be 180 degrees.  On a sphere, you can draw a triangle that contains a line along the equator, makes a right angle to the north pole, makes another right angle back to the equator..  Three right angles yields 270 degrees, which is possible on a sphere.  In 1998 a similar global geometric measurement became possible for our 3 dimensional universe, again by measuring a large triangle.  In this case,  the observational technique relied upon measurements of the Cosmic Microwave Background (CMB), the radiation `afterglow' of the big bang.  This radiation bath involves photons that last interacted when the universe was 1/1000th its present size, when it was also about 300,000 years old.  When we measure this background we are thus observing a surface from which radiation that has traveled almost 14 billion years to get to us.  Causality tells us that the largest lumps of matter that could even begin to collapse due to gravity are lumps that are no more than about 300,000 light years across.  Since gravity travels at the speed of light, lumps any larger than this do not even know they are lumps.  Now, the angular size of such 300,000 light-year-across lumps, as seen from the Earth will depend upon the propagation of the light rays that come from those lumps and make their way to our telescopes.  This propagation will be different depending upon the curvature of the intervening space.   In 1998, several ground based experiments, ultimately supplemented by the all sky microwave satellite called the Wilkenson Microwave Anisotropy Probe, were able to measure hot spots and cold spots in the CMB (corresponding to tiny excesses of matter and energy) and determine their spatial distribution and average angular size.  The results have been striking.  Within an accuracy of several percent they find (\cite{wmap})

\vskip 0.1in
\begin{equation}
\Omega_{tot} = 1.0 + 0.0084 - 0.0133 \ (95\%)
\end{equation}
\vskip 0.1in

{\bf {\it Supernovae}}.  The huge discrepancy between these two measurements of $\Omega$ suggest that there is another source of energy in the universe not associated with galaxies and clusters.  One possibility is the famous Cosmological Constant of Einstein.  While indirect evidence that a non-zero cosmological constant exists began to be suggested in earnest as early as 1995 (\cite{kraussturner}), a direct measurement of such a contribution occurred in 1998 when two separate groups reported evidence, based on observations of the luminosity distance to distant supernovae as a function of their redshift, that the expansion of the universe was accelerating! (\cite{perl,schmidt})  Such acceleration would result from a cosmological constant dominated universe, or any universe dominated by 'dark energy' with an equation of state similar to that of a cosmological constant, with $w ={p \over \rho} \approx -1$.   the Moreover, the contribution to $\Omega$ required to produce such acceleration is precisely consistent with the difference between the two measured values described above.     To date, all measurements of dark energy, determined by Supernova measurements, and by independent measurements of curvature, large scale structure, and the age of the universe are precisely consistent with a value of $w \approx -1$, namely a cosmological constant.

\vskip 0.2in

All cosmological measurements have now converged on a single, crazy cosmological model:  a flat universe, with approximately $70 \%$ of its energy in the form of something like a cosmological constant, $30 \%$ dark matter, and a few percent in matter corresponding to everything we can see with our telescopes.  This implies that the content of over $95\%$ of the univers is currently unknown.  Moreover, the discovery that the dominant energy in the universe resides in empty space changes absolutely everything about our understanding of the current and future behavior of the universe.  Textbooks used to state that when we determined the geometry of the universe we would be able to unambiguously predict the future.  A flat universe would expand forever, slowly asymtotically to a halt.  A closed universe would recollapse, and an open universe would continue expanding at a finite rate forever.   With the possible existence of dark energy, all of these predictions go out the window  (\cite{kraussturner2}).  An open universe can collapse and a closed universe can expand forever.  Until we know the nature of dark energy, whether or not it is truly a constant, we cannot predict the ultimate fate of the universe.  

The observation that the universe is flat vindicates the simplest prediction of so-called 'inflationary universe' scenarios of early universe physics.  But it also makes a more profound statement.  In a flat universe the total gravitational energy is precisely zero.  If the universe truly began from nothing, one would expect its total energy to be zero.  The fact that we observe this to be the case is therefore highly suggestive. 

The revolution that is currently taking place in our picture of cosmology because of the discovery of dark energy has been profound. While we do not have any good theoretical calculations that allow us to predict the observed value for the density of dark energy throughout space we currently understand that a cosmological constant could arise due to quantum mechanical effects combined with relativity, which imply that empty space is not actually empty but instead is full of  ` virtual' particle-antiparticle pairs that pop in and out the vacuum.   These virtual particles to contribute an energy to empty space resulting in a term that is identical to EinsteinÕs original cosmological term, leading to a universal repulsion and hence an accelerating universe.   

So far, so good.  However, when we attempt to estimate the magnitude of the vacuum energy on the basis of our current understanding of elementary particle physics, we get a value that is 120 orders of magnitude too large!
If the dark energy we observe corresponds to a cosmological constant arising from a non-zero vacuum energy in quantum mechanics, there is something fundamentally wrong with our fundamental theories of particle physics.    

This discrepancy is so great that some physicists are now resorting to appeals to the `anthropic principle' to explain it.  If the cosmological constant were much larger than it is, galaxies would never form in the early universe (assuming all other fundamental constants remain fixed), and if galaxies didn't form, then stars, planets and people wouldn't form.   So, if an infinity of universes exist in which the cosmological constant takes random values, we would only expect to find ourselves in universes in which the cosmological constant is not much larger than we actually observe!

As appealing as this picture may seem on the surface, it is rife with problems.  First of all, we do not know whether an infinite number of universe exist.  We also do not know the measure over all such possible universe.  We do not know if other constants of nature are also stochastic.  And finally, we have no idea if we are typical intelligent life forms.  Once all of these issues are examined, and our current state of ignorance is recognized, all predictive power associated with the anthropic principle vanishes. 

If anthropic arguments are correct (while we may never be able to falsify them), it suggests that science as we have known it for the past 400 years is, at a fundamental level, over.  The march of physics has constantly been aimed at showing that the universe must be the way it is, based on fundamental principles.  If anthropic arguments are correct, physics becomes, at a fundamental level merely an environmental science.

As depressing as this situation is, prospects for the long term future are even bleaker.   In a universe dominated by a cosmological constant, all observed structures outside of our own bound set of galaxies will eventually disappear.  The longer we wait, the less we will see, as distant objects recede from us at a faster and faster rate, with redshifts increasing exponentially until, in a period of about 1-10 trillion years, all evidence of their existence will cease. 

  Even as we have made remarkable progress in the past 30 years in unraveling a consistent, if perplexing picture of the universe, we must be wary of things we cannot now observe, and how they might alter that picture. Some perspective on our current intellectual situation may be gleaned by considering the physics and astronomy of the far future, based on the fact that information about distant galaxies will be lost, as a colleague and I have recently done (\cite{krausssh}). 
  
 In particular we find that in a time comparable to the age of the longest lived stars all evidence that the Big Bang will disappear--observers will not be able to perform any observation or experiment that infers either the existence of an
expanding universe dominated by a cosmological constant, or that there was a hot Big Bang.  

Why is this the case?  The three pillars of the big bang, are (1) the Hubble Expansion, (2) the existence of the CMB radiation, and (3) observations of the abundance of light elements.  As I have described, because of cosmic acceleration, in a period of perhaps 10 trillion years, comparable to the lifetime of the longest main sequence stars, all galaxies outside of our own bound cluster will disappear.  With no galaxies there will be no tracers of the Hubble expansion.  Interestingly, with no such tracers, all evidence of the existence of dark energy, and an accelerating universe will also disappear.  This will be true in spite of the fact that the dark energy will actually contribute a million million times more energy within the horizon than will visible matter at that time.  Yet it will be undetectable.  In fact, the only period when dark energy is detectable, if it has remained and continues to remain at its current value, is now.  At much earlier times it had a negligible effect upon the expansion, and at late times it drives out all tracers of the expansion. 

As far as the CMB is concerned, even if skeptical observers in the future
were inclined to undertake a search for this afterglow of the Big Bang, they would
come up empty-handed.  At $t \approx$ 100 Gyr, the peak
wavelength of the cosmic microwave background will be redshifted
to roughly $\lambda \approx 1$ m, or a frequency of roughly
300 MHz.  While a uniform radio background at this frequency would
in principle be observable, the intensity of the CMB will also
be redshifted by about 12 orders of magnitude.  At much later times, the CMB becomes
unobservable even in principle, as the peak wavelength is driven
to a length larger than the horizon.  Well before then, however,
the microwave background peak will redshift below
the plasma frequency of the interstellar medium, and so will be screened from any observer within the galaxy.
Recall that the plasma frequency is given by
$$\nu_p = \left(\frac{n_e e^2}{\pi m_e}\right)^{1/2},$$
where $n_e$ and $m_e$ are the electron number density and mass, respectively.  Observations of dispersion in pulsar
signals give \cite{ism} $n_e \approx 0.03$ cm$^{-3}$ in the interstellar medium, which corresponds to a plasma
frequency of $\nu_p \approx 1$ kHz, or a wavelength of $\lambda_p \approx 3 \times 10^7$ cm.  This corresponds
to an expansion factor $\sim 10^8$ relative to the present-day peak of the CMB.  Assuming an exponential expansion, dominated by dark energy, this expansion factor will be reached when the universe is less than 50 times its present age, well below the lifetime of the longest-lived main sequence stars.

The final key bit of evidence for the Big bang rests crucially
on the fact that relic abundances of deuterium remain observable at
the present day, while helium-4 has been enhanced by only a few percent
since it was produced in the early universe.  Extrapolating forward
by 100 Gyr, we expect significantly more contamination of the helium-4
abundance, and concomitant destruction of the relic deuterium.  It has been
argued that the ultimate extrapolation of light elemental abundances,
following many generations of stellar evolution, is a mass fraction of helium given by
$Y=0.6$.  The primordial
helium mass fraction of $Y=0.25$ will be a relatively small fraction of this
abundance.  It is unlikely that much deuterium could survive this degree of processing, and in any case the current ``smoking gun" of the deuterium abundance is provided by Lyman-$\alpha$
absorption systems, back-lit by QSOs.
Such systems will be unavailable to our observers of the future, as both the QSOs and the
Lyman-$\alpha$ systems will have redshifted outside of the horizon.

What will observers of the far future infer then?  Poetically their picture of the universe will not be significantly different than that which Einstein had when he developed general relativity: A static universe in which our galaxy was surrounded by eternal empty space, with which cosmology at the
turn of the last century began, will have returned with a vengeance.

We appear to live in a very special time: the only time when we can observationally verify that we live at a very special time!  Of course, however, it is likely that all observers at all times may feel this is the case.  What considerations of the future tell is that we perhaps should have some cosmic humility. Perhaps observations made in the far future will reveal aspects of the universe we cannot yet measure that will change the current, strange model we have uncovered for the cosmos, or that will reveal aspects of dark energy that we cannot now even imagine. 

The last century has clearly changed our understanding of the place of humanity, and the earth, within the cosmos.  The revolutions of the past several decades have changed everything about our understanding of the future.  Will these revolutions change our cultural perspective in a way that changes behavior?   Will the current mysteries surround dark energy provide ultimate limits on the way we carry out fundamental science?  Do we really live at a special time in the history of the universe?  Time will tell, but I am encouraged by the fact  that nature keeps surprising us in ways that are well beyond our human imagination, which is why science can, and should, continue to alter our cultural perceptions, guiding humanity to better appreciate the remarkably fortunate circumstances under which we are able to explore the world around us.

\vskip 0.1in
\noindent{This work was supported in part by a grant from the U.S. Department of Energy and by startup funding from Arizona State University}

\end{document}